\documentclass[onecolumn,11pt,a4paper,notitlepage]{article}

\usepackage{amssymb}
\usepackage{epsfig}

\def\tfrac#1#2{{\textstyle {#1 \over #2}}}%
\begin{document}

\title{Sensitivity, Itinerancy and Chaos in Partly--Synchronized Weighted
Networks}
\author{{\small J. Marro}$^{+}${\small , J.J. Torres}$^{+},${\small \ J.M.
Cortes}\thanks{%
Presently at Institute for Adaptive and Neural Computation, School of
Informatics, University of Edinburgh, EH1 2QL, UK.}$\ ^{+\S }${\small , and
B. Wemmenhove}$^{\S }$ \\
$^{+}${\small Institute Carlos I for Theoretical and Computational Physics,
and}\\
{\small \ Departamento de Electromagnetismo y F\'{\i}sica de la Materia,}\\
{\small \ University of Granada, E-18071 Granada, Spain.}\\
{\small \ }$^{\S }${\small Department of Biophysics and SNN, Radboud
University Nijmegen,}\\
{\small \ 6525 EZ Nijmegen, The Netherlands}}
\maketitle

\begin{abstract}
We present exact results, as well as some illustrative Monte Carlo
simulations, concerning a stochastic network with weighted connections in
which the fraction of nodes that are dynamically synchronized, $\rho \in %
\left[ 0,1\right] ,$ is a parameter. This allows one to describe from
single--node kinetics $(\rho \rightarrow 0)$ to simultaneous updating of all
the variables at each time unit $(\rho \rightarrow 1).$ An example of the
former limit is the well--known sequential updating of spins in kinetic
magnetic models whereas the latter limit is common for updating complex
cellular automata. The emergent behavior changes dramatically as $\rho $ is
varied. For small values of $\rho ,$ we observe relaxation towards one of
the attractors and a great sensibility to external stimuli and, for $\rho
\geq \rho _{c},$ itinerancy as in heteroclinic paths among attractors;
tuning $\rho $ in this regime, the oscillations with time may abruptly
change from regular to chaotic and vice versa. We show how these
observations, which may be relevant concerning computational strategies,
closely resemble some actual situations related to both searching and states
of attention in the brain.\smallskip

\end{abstract}

\section{Introduction and definition of basic model}

A dynamic network of many nodes connected by weighted communication lines
models a great variety of situations in physics, biology, chemistry and
sociology, and it has a wide range of technological applications as well;
see, for instance, \cite%
{CNN1,marroB,newmannets,weight1,cp2005,nets05,weight2}. Examples of weighted
networks are the metabolic and food webs, which connect chains of different
intensity, the Internet, the world wide web and other social networks, in
which agents may interchange different amounts of information or money, the
transport nets, whose connections differ in capacity, number of transits
and/or passengers, spin glasses and reaction--diffusion systems, in which
diffusion, local rearrangements and reactions may vary the effective ionic
interactions, and the immune system, the central nervous system and the
brain, e.g., high--level functions in the latter case seem to rely on
synaptic changes.

A rather general feature in these systems is that the nodes are not fully
synchronized when performing a given task ---which may be either a matter of
economy or else, perhaps more frequently, a necessary condition for
efficient performance. Even though this is as evident as the fact that links
are seldom homogeneous, studies of partly synchronized networks are rare.
Furthermore, the relevant literature is dispersed, as it was generated in
various distant fields, and a broad coherent description is lacking. In
particular, related studies often disregard an important general property,
namely, that the systems of interest are out of equilibrium. That is, they
cannot settle down into an equilibrium state but the network typically keeps
wandering in a complex space of fixed points or, in one of the simplest
cases, it reaches a \textit{nonequilibrium} steady state whose nature
depends on dynamic details \cite{marroB}. This results in a complex
landscape of emergent properties whose relation with the network details is
poorly understood. In this paper, as a new effort aimed at methodizing
somewhat the picture, we present some related exact results, together with
illustrative Monte Carlo simulations, which apply to a rather general class
of partly--synchronized heterogeneous or weighted networks. It follows, as a
first application, examples of itinerancy and constructive chaos which mimic
recent experimental observations.

Consider a network, with a processor, neuron, spin or, simply, variable at
each node, and define the sets of node activities, $\mathbf{\sigma }\equiv
\left\{ \sigma _{i}\right\} ,$ and communication line weights, $\mathbf{w}%
\equiv \left\{ w_{ij}\right\} ,$ where $i,j=1,\ldots ,N.$ Each node is acted
on by a local field $h_{i}\left( \mathbf{\sigma },\mathbf{w}\right) $ which
is induced by the weighted action of the other, $N-1$ nodes. We also define
an additional, operational set of binary indexes, $\mathbf{x}=\{ x_{i}=0 
\,\, \mathrm{or} \,\, 1 \} .$ Time evolution proceeds according to a generalized
cellular--automaton strategy. That is, at each time unit, one simultaneously
modifies the activity of $n$ variables, $1\leqslant n\leqslant N,$ and the
probability of the network state evolves in discrete time, $t,$ according to 
\begin{equation}
P_{t+1}(\mathbf{\sigma })=\sum_{\mathbf{\sigma }^{^{\prime }}}R\left( 
\mathbf{\sigma }^{\prime }\mathbf{\rightarrow }\mathbf{\sigma }\right) P_{t}(%
\mathbf{\sigma }^{\prime })  \label{meq}
\end{equation}%
with the (microscopic) transition rate:%
\begin{equation}
R\left( \mathbf{\sigma }\mathbf{\rightarrow }\mathbf{\sigma }^{\prime
}\right) =\sum_{\mathbf{x}}p_{n}(\mathbf{x})\prod_{\left\{ i|x_{i}=1\right\}
}\tilde{\varphi}_{n}\left( \sigma _{i}\rightarrow \sigma _{i}^{\prime
}\right) \prod_{\left\{ i|x_{i}=0\right\} }\delta _{\sigma _{i},\sigma
_{i}^{\prime }}.  \label{rate}
\end{equation}%
Here, $\tilde{\varphi}_{n}\left( \sigma _{i}\rightarrow \sigma _{i}^{\prime
}\right) \equiv \varphi \left( \sigma _{i}\rightarrow \sigma _{i}^{\prime
}\right) \left[ 1+\left( \delta _{\sigma _{i}^{\prime },-\sigma
_{i}}-1\right) \delta _{n,1}\right] $ is the elementary local rate, $\varphi 
$ is an arbitrary function of $\beta \sigma _{i}h_{i}$ where $\beta $ is an
inverse \textquotedblleft temperature\textquotedblright\ ---i.e., a
parameter which controls the stochasticity of the process--- and, for any
set of $n$ sites chosen at random, one has that%
\begin{equation}
p_{n}\left( \mathbf{x}\right) =\left( 
\begin{array}{c}
N \\ 
n%
\end{array}%
\right) ^{-1}\delta \left( \sum_{i}x_{i}-n\right) .  \label{px}
\end{equation}

This is the natural generalization of two familiar cases: The Glauber
sequential updating \cite{marroB} follows for $n=1,$ so that it is obtained
macroscopically in the limit of the minimal dynamic perturbation or $\rho
\equiv n/N\rightarrow 0,$ while the above reduces to the Little parallel
updating \cite{little,CA3} for $n=N,$ i.e., as $\rho \rightarrow 1.$ One may
think of many situations whose understanding may benefit from studying the
crossover between these two situations. For example, assuming a cell which
is stimulated only in the presence of a neuromodulator such as dopamine, the
parameter $n$ will correspond to the number of neurons that are modulated
each cycle. That is, the other $N-n$ neurons receive no input but maintain
memory of the previous state, 
which has been claimed to be at the basis of working memories \cite{micro}.

\section{Some explicit realizations}

For simplicity, and also to have well defined references, we shall represent
nodes in the following as binary variables, $\sigma _{i}=\pm 1,$ while $%
w_{ij}\in 
\mathbb{R}
,$ and consider local rates $\varphi \left( \sigma _{i}\rightarrow \sigma
_{i}^{\prime }=-\sigma _{i}\right) =%
\frac{1}{2}%
\left[ 1-\sigma _{i}\tanh \left( \beta h_{i}\right) \right] ,$ which is
rather customary as a case that satisfies detailed balance \cite{marroB}.
Notice, however, that, in general, detailed balance is not fulfilled by our
basic equation (\ref{meq}) nor by the superposition (\ref{rate}) as far as $%
n>1.$ Consequently, in general, our system cannot be described by Gibbs
ensemble theory. We shall further assume that the fields satisfy 
\begin{equation}
h_{i}\left( \mathbf{\sigma },\mathbf{w}\right) =h\left[ \mathbf{\pi }\left( 
\mathbf{\sigma }\right) ,\mathbf{\xi }_{i}\right] .  \label{hi}
\end{equation}%
We are assuming here a set of $M$ given \textit{patterns}, namely, different
realizations of the network set of activities, to be denoted as $\mathbf{\xi 
}\equiv \left\{ \mathbf{\xi }_{i}\right\} $ with $\mathbf{\xi }_{i}\equiv
\left\{ \xi _{i}^{\mu }=\pm 1;\mu =1,\ldots ,M\right\} ,$ and $\mathbf{\pi }%
\equiv \left\{ \pi ^{\mu }\left( \mathbf{\sigma }\right) \right\} ,$ where
the product $\pi ^{\mu }\left( \mathbf{\sigma }\right) =N^{-1}\sum_{i}\xi
_{i}^{\mu }\sigma _{i}$ measures the \textit{overlap} of the current state
with pattern $\mu .$ For $N\rightarrow \infty $ and finite $M,$ i.e., in the
limit $\alpha \equiv M/N\rightarrow 0,$ from (\ref{meq})--(\ref{hi}), the
mesoscopic time--evolution equation%
\begin{equation}
\pi _{t+1}^{\mu }\left( \mathbf{\sigma }\right) =\rho N^{-1}\sum_{i}\xi
_{i}^{\mu }\tanh \left\{ \beta h_{i}\left[ \mathbf{\pi }_{t}\left( \mathbf{%
\sigma }\right) ,\mathbf{\xi }_{i}\right] \right\} +\left( 1-\rho \right)
\pi _{t}^{\mu }\left( \mathbf{\sigma }\right)   \label{mt}
\end{equation}%
follows for any $\mu .$ The details of the derivation, as well as some
possible generalizations of this result, will be published elsewhere \cite%
{CortesNew}.

One of the simplest realizations of the above is the Hopfield network \cite%
{hopf1,hopf2,perettoB}. In this case, the communication line weights are
heterogeneous but fixed according to the Hebb (\textit{learning})
prescription $w_{ij}=N^{-1}\sum_{\mu }\xi _{i}^{\mu }\xi _{j}^{\mu },$ and
the local fields are $h_{i}\left( \mathbf{\sigma },\mathbf{w}\right)
=\sum_{j\neq i}w_{ij}\sigma _{j}.$ These choices satisfy condition (\ref{hi}%
) which also holds for other non--linear \textit{learning} rules; in any
case, one may easily generalize (\ref{mt}) to include other interesting
cases (see, for instance, \cite{learn}) which do not precisely conform to (%
\ref{hi}). The original Hopfield model evolves by Glauber processes, namely,
by attempting a single variable change, $\sigma _{i}\rightarrow -\sigma
_{i}, $ at each unit time ---e.g., the Monte Carlo step--- with probability $%
\varphi \left( \beta \sigma _{i}h_{i}\right) .$ The symmetry $w_{ij}=w_{ji}$
and detailed balance then guarantee asymptotic convergence to \textit{%
equilibrium}, i.e., $P_{t\rightarrow \infty }\left( \mathbf{\sigma }\right)
\propto \exp \left( \beta \sum_{i}h_{i}\sigma _{i}\right) .$

Computational efficiency has sometimes motivated to induce time evolution of
the Hopfield network by the Little strategy, i.e., $\rho \rightarrow 1$ in
our formulation. This is known to drive the system to a full \textit{%
nonequilibrium} situation, in general \cite{grin}. The local rule and other
details of dynamics are then essential in determining the emergent behavior.
Not only the time evolution may vary but also the nature of the resulting
asymptotic state, perhaps including morphology, phase diagram, universality
class, etc.; see Refs.\cite{marroB,odorRMP,mdiezPRL} for some outstanding
examples of this assertion.

For completeness, we mention that the Hopfield network, i.e., $\rho
\rightarrow 0$ will also correspond, in general, to a nonequilibrium
situation when implemented (unlike in the original proposal \cite{hopf1})
with asymmetric or time--evolving weights or with a dynamic rule lacking
detailed balance. There is some chance that an \textit{effective Hamiltonian 
}\cite{GMeff}\textit{, }such that $P_{t\rightarrow \infty }\left( \mathbf{%
\sigma }\right) \propto \exp \left( H_{\mathrm{eff}}\right) ,$ can then be
defined, however. When this is the case, one may often apply equilibrium
methods, with the result of relatively simple emergent properties \cite%
{torresJPA,cortesNC}.

Concerning our proposal (\ref{mt}), we first mention that, assuming fields
that conform to the Hebb prescription with static weights, the Hopfield
property of \textit{associative memory }is recovered for $\rho \rightarrow
0, $ as expected. That is, for high enough $\beta $ (which means below
certain stochasticity) the patterns $\mathbf{\xi }$ may be attractors of
dynamics. Consequently, an initial state resembling one of the patterns,
e.g., a degraded picture will converge towards the original one, which
mimics simple recognition \cite{hopf2,perettoB}. We checked too that, in
agreement with some previous indications \cite{herzPRE}, implementing the
Hopfield--Hebb network with $\rho >0$ produces the behavior that
characterizes the familiar case $\rho \rightarrow 0,$ including associative
memory, even though equilibrium is precluded, e.g., in general, no effective
Hamiltonian is predicted to exist for any $\rho >0$ \cite{grin}. Excluding
these Hopfield--Hebb versions, our model exhibits a complex behavior which
depends dramatically on the value of $\rho .$ This is a consequence of
changes with $\rho $ in the stability associated with (\ref{mt}), as we show
next.

\section{Some main results}

The local fields may be determined according to various criteria, depending
on the specific application of interest. That is, one may investigate the
consequences of equation (\ref{mt}) and associated stability for different
relations between the fields and the weights $w_{ij}$ and between these and
other network properties.

\begin{figure}
\centerline{\psfig{file=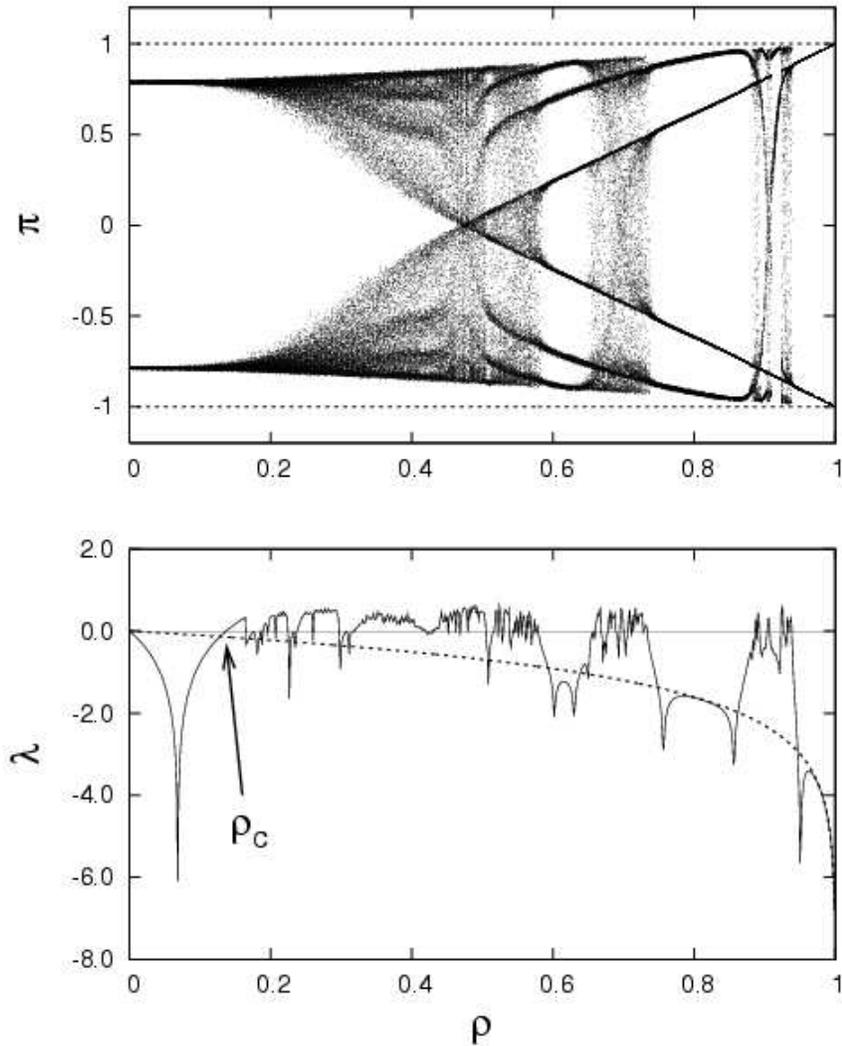,width=12cm} }
\caption{Evidence for critical
transitions from regular to chaotic behavior as the \textit{synchronization
parameter} $\rho =n/N$ is varied. \underline{Top graph:}
Monte Carlo results for depressing fluctuations with $\Phi =1/2.$ This shows
the dependence on $\rho $ of the stationary overlap between the
activity and a single (randomly generated) pattern for $N=3600$ variables,
and inverse \textquotedblleft temperature\textquotedblright\ $\beta %
=20.$ The dashed line is for the standard Hopfield--Hebb case, $\Phi =-1.$ 
\underline{Bottom graph:} The dependence on $\rho $ of the
Lyapunov exponent, as obtained analytically from the saddle--point solution,
for the same case as above (solid irregular line) and for $\Phi =-1$ (dashed
line). This graph also shows the value $\rho =\rho _{c},$
i.e., the minimum degree of synchronization needed to observe chaotic
behavior, and the line $\lambda =0$ for reference purposses.
}
\label{figure1}
\end{figure}

We shall be mostly concerned in the rest of this paper with a specific 
\textit{neural automaton} as a working example. In this case, the above
assumption of static line weights happens to be rather unrealistic. As a
matter of fact, one is eager to admit, concerning different contexts, that
the communication line weights may change with the nodes activity, and even
that they may loose some competence after a time interval of heavy work.
This seems confirmed in the case of the brain where the transmission of
information and many computations are strongly correlated with
activity--induced fast fluctuations of the synaptic intensities, namely, our 
$w_{ij}$'s \cite{noise,abb}. Furthermore, assuming the experimental
observation that synaptic changes may induce \textit{depression} \cite%
{depre0} seems to have important consequences \cite{depre1,bibit,cortesNC}.
That is, a repeated presynaptic activation may decrease the neurotransmitter
release, which will depress the postsynaptic response and, in turn, affect
noticeably the system behavior. For concreteness, motivated by these facts,
we shall adopt here the proposal in Refs.\cite{cortesNC,marroPRE}. This
amounts to assume a simple generalization of the Hebb prescription which is
in accordance with condition (\ref{hi}), namely,%
\begin{equation}
w_{ij}=\left[ 1-\left( 1+\Phi \right) q\left( \mathbf{\pi }\right) \right]
N^{-1}\sum_{\mu =1}^{M}\xi _{i}^{\mu }\xi _{j}^{\mu },  \label{wnew}
\end{equation}%
where $q\left( \mathbf{\pi }\right) \equiv \left( 1+\alpha \right) \sum_{\mu
}\pi ^{\mu }\left( \mathbf{\sigma }\right) ^{2}$ depends on the set of
stored patterns. The Hopfield case discussed above is recovered for $\Phi
=-1,$ while other values of this parameter correspond to fluctuations which
induce depression of synapses by a factor $-\Phi $ on the average.

The choice (\ref{wnew}) happens to importantly modify the network behavior,
even for a single \textit{stored }pattern, $M=1.$ The stationary, $%
t\rightarrow \infty $ solution of (\ref{mt}) is then $\pi _{\infty }=F\left(
\pi _{\infty };\rho ,\Phi \right) $ with 
\begin{equation}
F\left( \pi ;\rho ,\Phi \right) \equiv \rho \tanh \left\{ \beta \pi \left[
1-\left( 1+\Phi \right) \pi ^{2}\right] \right\} +\left( 1-\rho \right) \pi ,
\label{efe}
\end{equation}%
and local stability requires that $\left\vert \partial F\left( \pi ;\rho
,\Phi \right) /\partial \pi \right\vert <1.$ The fixed point is therefore $%
\pi _{\infty }=\tanh \left\{ \beta \pi _{\infty }\left[ 1-\left( 1+\Phi
\right) \pi _{\infty }^{2}\right] \right\} ,$ independent of $\rho ,$ while
stability crucially depends on $\rho .$ The limiting condition $\left.
\partial F\left( \pi ;\rho ,\Phi \right) /\partial \pi \right\vert _{\pi
_{\infty }}=1$ corresponds to a steady--state 
bifurcation. This implies for $\pi _{\infty }=0$ that $\beta <\beta _{c}=1,$
independent of both $\rho $ and $\Phi .$ Non--trivial solutions $\pi
_{\infty }\neq 0$ in this case require that $\Phi >-4/3,$ which includes the
Hopfield case. The other limiting condition $\left. \partial F\left( \pi
;\rho ,\Phi \right) /\partial \pi \right\vert _{\pi _{\infty }}=-1$
corresponds to a period doubling bifurcation. It follows from this that
local stability requires $\rho <\rho _{c}$ with%
\begin{equation}
\rho _{c}=2\left\{ 3\beta \pi _{\infty }^{2}\left[ \left( \Phi +\tfrac{4}{3}%
\right) -\left( 1+\Phi \right) \pi _{\infty }^{2}\right] -\beta +1\right\}
^{-1}.  \label{pc}
\end{equation}%
It is to be remarked that this condition cannot be fulfilled in the
Hopfield, $\Phi =-1$ case, for which one obtains from (\ref{pc}) the
nonsense solution $\rho _{c}\geq 2.$

\begin{figure}
\centerline{\psfig{file=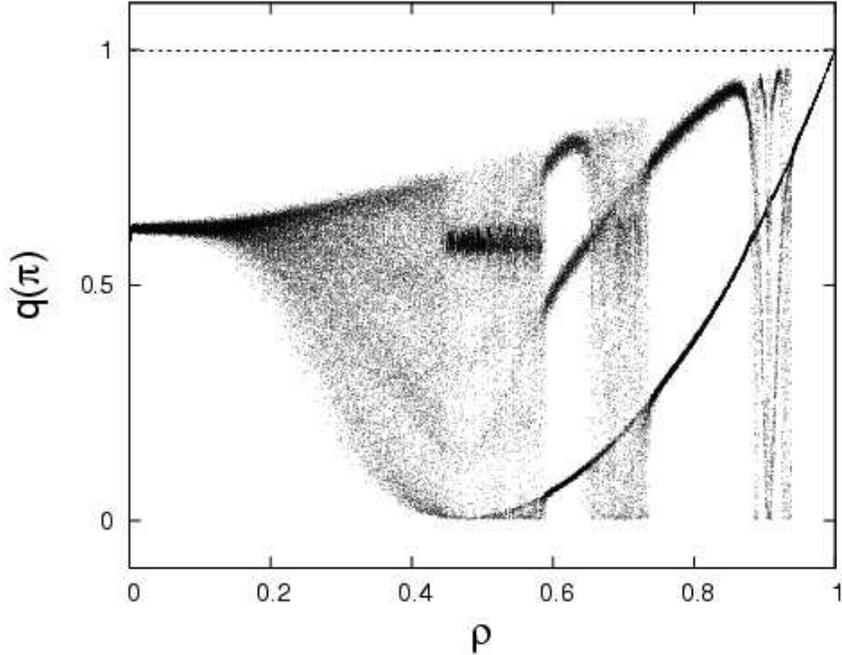,width=12cm} }
\caption{The chaotic behavior in figure 
\ref{figure1} occurs also as the number of patterns is increased.
This graph shows the dependence on the synchronization parameter $%
\rho $ of the steady--state value of the order parameter $q\left( \mathbf{%
\pi }\right) \equiv \left( 1+\alpha \right) \sum_{%
\mu }\pi ^{\mu }\left( \mathbf{\sigma }\right) ^{2}.$
The behavior shown here ---which follows indistinctly from the analytical
solution or from Monte Carlo simulations--- is for $M=20$ stored (randomly
generated) patterns, $N=3600,$ $\beta =20$ and $\Phi =1/2.$ The
dashed, horizontal line $q=1$ is for the Hopfield case, $\Phi =-1;$ this
illustrates that, except for scaling of the relaxation time, the value of $%
\rho $ is then irrelevant.
}
\label{figure2}
\end{figure}

The resulting behavior is illustrated in figure \ref{figure1}. This shows,
for $M=1,$ the onset of chaos at $\rho =\rho _{c}$ in the saddle--point map (%
\ref{mt}) and, accurately fitting this, in Monte Carlo simulations. The
behavior shown in the top graph of figure \ref{figure1}, which is for $\Phi =%
\frac{1}{2}%
,$ is likely to characterize any $\Phi \neq -1$ as well. This behavior does
not occur for the singular Hopfield case with static synapses, for which the
stability of $\mathbf{\pi }$ is independent of $\rho .$ The bottom graph in
figure \ref{figure1} illustrates that one has for $\rho >\rho _{c}$ regimes
of regular oscillations among the attractors (i.e., the given pattern and
its \textquotedblleft negative\textquotedblright\ in this case with $M=1)$
which are eventually interrupted as one varies $\rho ,$ even slightly, by
eventual chaotic jumping.

The critical value of the synchronization parameter for the emergence of
chaos, due to local instabilities around the steady $\pi $ solution may be
estimated in figure \ref{figure1} as $\rho _{c}=0.137$ for $M=1.$ This is
precisely the value that one obtains from equation (\ref{pc}) for $\beta =20$%
, $\pi =0.788$ and $\Phi =%
\frac{1}{2}%
.$

\begin{figure}
\centerline{\psfig{file=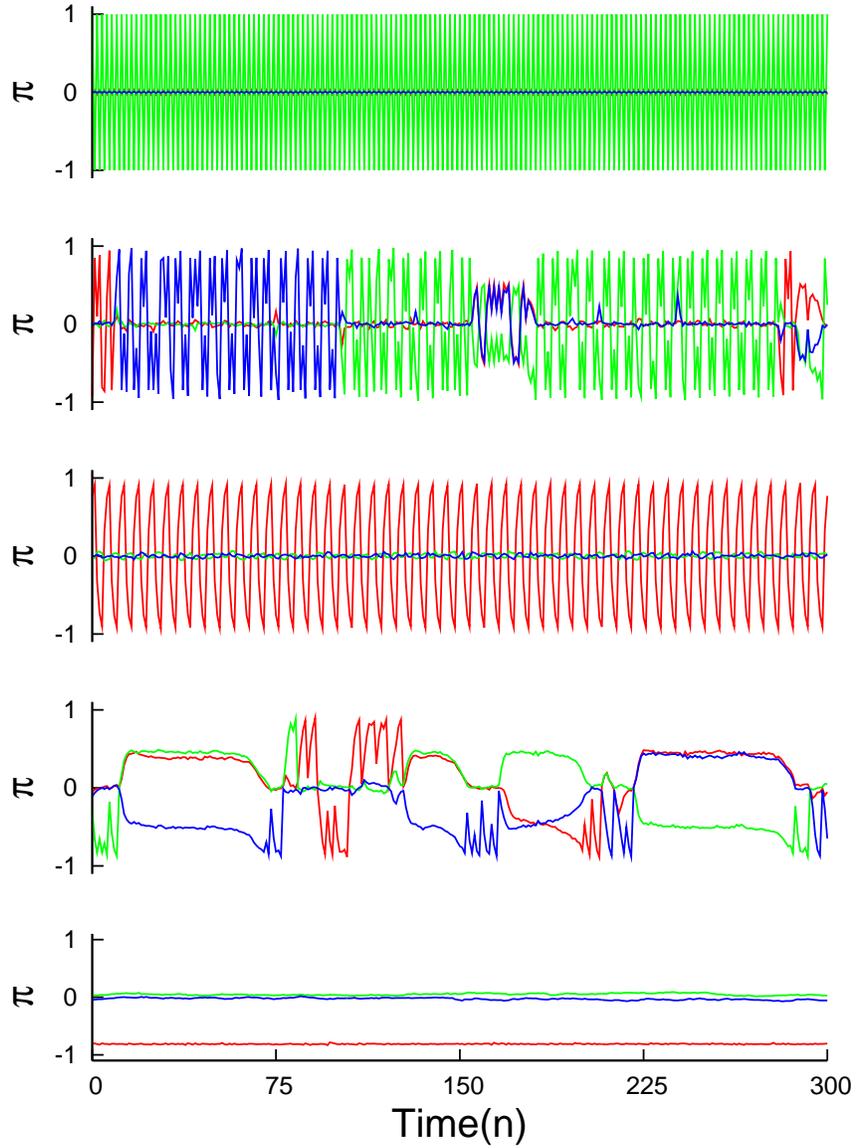,width=12cm} }
\caption{The overlap as a function of
time (in units of $n$ MC trials) after a transient time $t=1920n,$ for $%
N=1600,$ $\beta =20,$ $\Phi =0.4$ and $M=3$ uncorrelated patterns.
It follows in this case that $\rho _{c}=0.085.$ Different graphs are
for increasing values of $\rho $ from bottom to top, namely, for $%
\rho =0.08,$ 0.50, 0.65, 0.92 and 1.00, respectively.
}
\label{figure3}
\end{figure}

Figure \ref{figure2} confirms that this behavior occurs also for $M>>1$
stored patterns, and figure \ref{figure3} shows the detail during the 
\textit{stationary} part of typical runs for representative values of $\rho
. $ In particular, figure \ref{figure3} illustrates qualitatively--different
types of time series our model exhibits, namely, from bottom to top: (%
\textit{i}) convergence towards one of the attractors ---in fact, to one of
the \textit{antipatterns}--- for $\rho <\rho _{c};$ (\textit{ii}) chaos,
i.e., fully irregular behavior with a positive Lyapunov exponent for $\rho
>\rho _{c};$ (\textit{iii}) a perfectly regular oscillation between one of
the attractors and its negative for $\rho >\rho _{c};$ (\textit{iv}) onset
of chaotic oscillations as $\rho $ is increased somewhat; and (\textit{v})
very rapid and completely ordered and periodic oscillations between one
pattern and its antipattern when all the nodes evolve synchronized with each
other at each time step. The cases (\textit{ii}) and, less markedly, (%
\textit{iv}) are nice examples of instability--induced switching phenomena.
That is, as suggested also in experiments on biological systems (see next
section), the network describes heteroclinic paths among the stored
patterns, remaining for different time intervals in the neighborhood of
different attractors, the choice of attractor being at random.

\section{Discussion\label{secDisc}}

In summary, we report in this paper on a class of homogeneous or weighted
networks in which the density, $\rho ,$ or the number of variables that are
synchronously updated at a time may be varied. This is remotely related to 
\textit{block--dynamics,} \textit{block--sequential, }and associated
algorithms \cite{randall1,randall2}, which aim at more efficient
computations, and it generalizes some previous proposals \cite%
{rossPRA,herzPRE,Jcortes}. We describe in detail the behavior of a
particular realization of the class, namely, a \textit{neural automaton}
which is motivated by recent neurobiological experiments and related
theoretical analysis. Different realizations of the class correspond to
different choices of the local fields $h\left( \mathbf{\pi },\mathbf{\xi }%
\right) $ that act upon the stochastic variables at the (neural) nodes. For
certain values of these fields, which amount to fix the (synaptic)
connections at some constant values, e.g., according to the Hebb
prescription, one recovers the equilibrium Hopfield network. The parameter $%
\rho $ is then irrelevant concerning most of the system properties. Our
model also admits simple extensions, corresponding to other values of the
local fields, that one may characterize by a complex \textit{effective
temperature} \cite{marroB,cortesNC}.

\begin{figure}
\centerline{\psfig{file=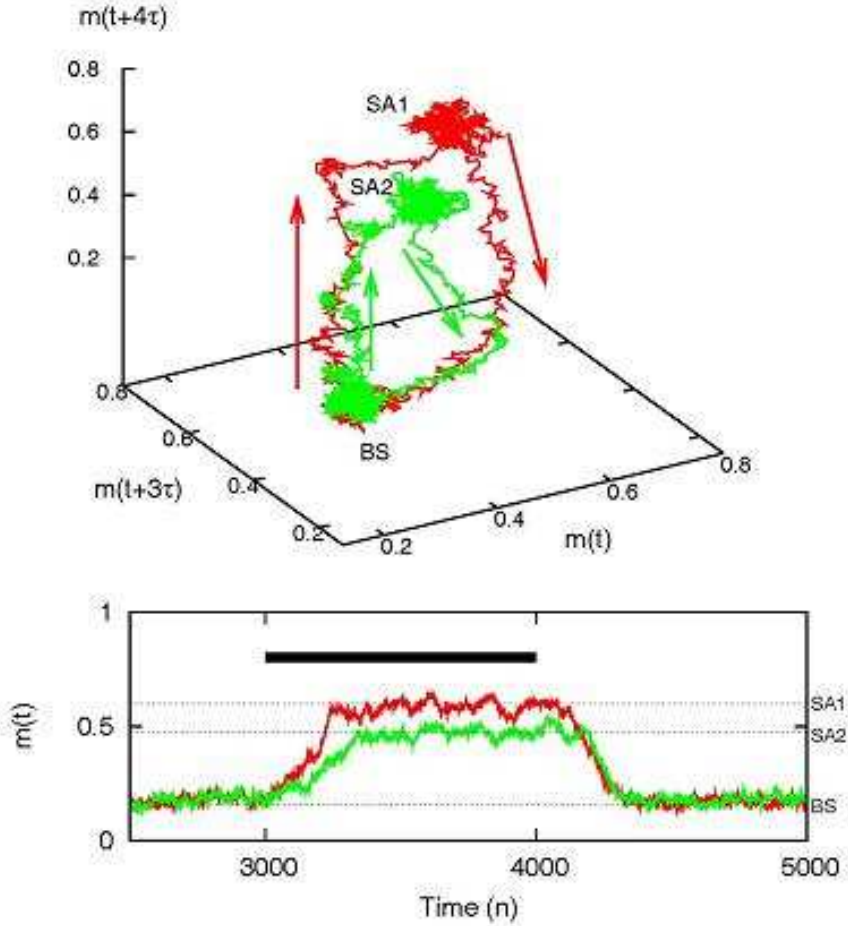,width=12cm} }
\caption{Itinerancy induced by an
external stimulus. This shows the mean firing rate as a function of time
(bottom graph) and as a \textit{phase space} three--dimensional trajectory
(top graph) when the system is perturbed as explained in the main text. The
situation here closely resembles the experimental observation concerning
odor responses in figures 2 and 4 in Ref. \cite{mazor}. The graphs
here correspond to the Monte Carlo simulation of our system with $N=1600,$ $%
T=0.25,$ $\Phi =0.45,$ $\protect\rho =3/64<\protect\rho _{c},$ and six
stored patterns. The time unit corresponds to $n$ trials. Performing the top
graph required a standard false--neighbor method which indicated an \textit{%
embedding dimension} \cite{embed} of 5. The involved time delay is $%
\tau =20n$ MC trials.
}
\label{figure4}
\end{figure}

The Hopfield picture has severe limitations concerning its practical
usefulness \cite{perettoB}, and some of these limitations may be overcome by
producing a full nonequilibrium condition. It is sensible to expect that $%
\rho $ will then transform into a relevant parameter. This is, in fact, the
situation in \cite{rossPRA} which deals with a modification of the Hebb
prescription which includes multiple interactions, random dilution and a
Gaussian noise. This implies a choice for the fields that even precludes the
existence of an effective temperature, and chaotic neural activity ensues.
Our biologically motivated choice, namely, Hopfield local fields with the
simple prescription (\ref{wnew}), which is interpreted as a consequence of
depressing synaptic fluctuations, induces a full nonequilibrium condition
---even for $\rho \rightarrow 0.$ In this limit, the system was recently
shown to exhibit enhancement of the network sensitivity to external stimuli 
\cite{cortesNC}. This happens to be a feature of the system for any $\rho .$
This interesting behavior, which we illustrate in figure \ref{figure4},
corresponds to a type of instability which is known to occur in nature, as
discussed below. This is associated here to a modification of the topology
of the space of fixed points due to the action of the involved $\Phi $%
--controlled noise.

Figure \ref{figure4} is a model remake of experiments on the odor response
of the (projection) neurons in the locust antennal lobe \cite{mazor}. Our
simulation illustrates two time series (with different colors) for the mean
firing rate, $m=\frac{1}{2N}\sum_{i}\left( 1+\sigma _{i}\right) ,$ in a
system with six stored patterns which is exposed to two different stimuli of
the same intensity and duration (between 3000 and 4000 time steps ---each
step corresponding here to $n$ trials). Each pattern consists of a string of 
$N$ binary variables; three of them are generated at random with,
respectively, 40\%, 50\% and 60\% of the variables set equal to 1 (the rest
are set equal to $-1),$ while the other three have the 1s at the first 70\%,
50\% and 20\% positions in the string, respectively. The bottom graph shows
with horizontal lines the baseline activity without stimulus (BS) and the
network activity level in the presence of the stimulus $\mu =1$ (SA1) and $%
\mu =2$ (SA2), which correspond to two of the random stored patterns.

The conclusion is that the stimulus destabilizes the system activity as in
the laboratory experiments. Note, however, that this occurs for $\rho <\rho
_{c},$ i.e., in the absence of chaotic behavior. As a matter of fact, the
behavior in figure \ref{figure4} will also be exhibited in the limit $\rho
\rightarrow 0$ which does not show any irregular behavior \cite{cortesNC}.

On the other hand, the suggestion that fluctuating connections may induce
fractal or \textit{strange} attractors as far as $\rho >0$ \cite%
{Jcortes,marroPRE} is confirmed here. As illustrated in figure \ref{figure3}%
, our system may exhibit both static and kind of dynamic associative memory
in this case. That is, the network state either will go to one of the
attractors (corresponding to one of the given patterns \textit{stored} in
the connecting synapses) or else, for $\rho \geq \rho _{c},$ will forever
remain visiting several or perhaps all the possible attractors. Furthermore,
the inspection rounds may abruptly become irregular and even chaotic as the
density of synchronized neurons varies slightly. It follows, in particular,
that the most interesting, oscillatory behavior requires synchronization of
a minimum of variables, and also that occurring chaotic jumps between
attractors requires some careful tuning of $\rho .$ In fact, as illustrated
in the bottom graph of figure \ref{figure1}, once the critical condition $%
\rho >\rho _{c}$ is fulfilled a complex situation ensues where it seems
difficult to predict the resulting behavior for slight variations of $\rho .$

\begin{figure}
\centerline{\psfig{file=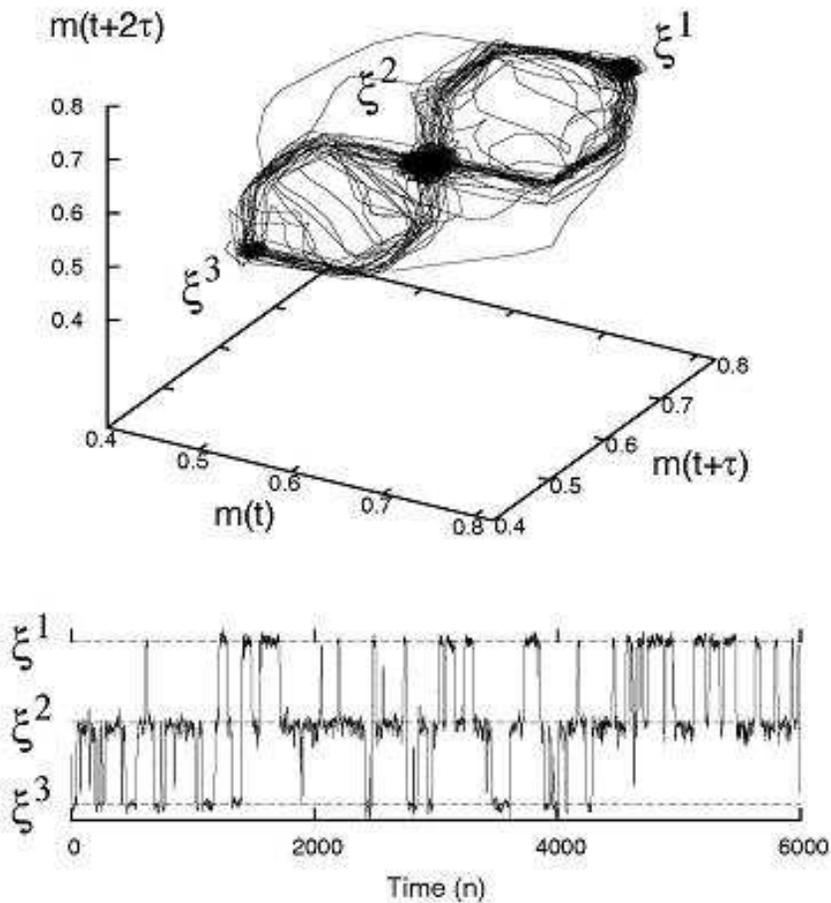,width=12cm} }
\caption{Instability and switching among
the attractors as in a state of attention induced by chaos. The mean firing
rate as a function of time (bottom graph) and as a trajectory in a
three--dimensional \textit{phase space} (top graph) correspond here to a
simulation with $N=1600,$ $T=0.006,$ $\Phi =%
\frac{1}{2},$ $\rho =123/320>\rho _{c}$ in a chaotic window, and three
stored patterns denoted $\xi ^{\mu }.$ The top graph
involves an embedded dimension of 5 and a time delay $\tau =5n$ MC
trials.
}
\label{figure5}
\end{figure}

The chaotic behavior is further illustrated in figure \ref{figure5}. This
shows trajectories among the three stored patterns, namely, $\mathbf{\xi }%
^{1},$ $\mathbf{\xi }^{2}$ and $\mathbf{\xi }^{3},$ $\mathbf{\xi }^{\mu
}=\left\{ \xi _{i}^{\mu };i=1,\ldots ,N\right\} .$ These are designed,
respectively, as a homogeneous string of 1s, a string with the first 50\%
positions set to 1 and the rest to $-1,$ and as a string with only the first
20\% positions set to 1. We observe many jumps between two close (more
correlated) patterns and, eventually, a jump to the most distant pattern.

It seems sensible to comment on this behavior at the light of the growing
evidence of chaotic behavior in the nervous system \cite{caos1,caos2}. We
have shown (e.g., figure \ref{figure4}) that chaos is not needed to have
efficient adaptation to a changing environment. However, one may argue \cite%
{caos1}, for instance, that the instability inherent to chaotic motions
facilitates this and, in particular, the system ability to move to any
pattern at any time. This behavior, which has been described for the
activity of the olfactory bulb and other cases, is nicely illustrated in
figures \ref{figure3} and \ref{figure5}. Our network thus mimics the
observed correlation between chaotic neuron activity and states of attention
in the brain, as well as other cases of constructive chaos in biology \cite%
{attent1,attent2,attent2bis,attent3}.

Chaos has been reported in other interesting networks (e.g., Ref. \cite%
{domin}) but, to our knowledge, never in such a general setting as here. As
a matter of fact, the present model allows for some natural generalizations 
\cite{CortesNew} and, in particular, suggests a great interest for a more
detailed study of the apparently unpredictable behavior it exhibits for $%
\rho >\rho _{c}.$ This teaches us that varying $\rho $ is a simple method to
control chaos in networks, and that this may also help in determining
efficient computation strategies \cite{edge1,edge2}. Concerning the latter,
the model behavior may be relevant, for instance, when judging on the best
procedure for specific data mining and for the control of different
activities on a multiprocessor system \cite{muellerDM}, and deciding on
whether to implement sequential or parallel programming in some extreme
cases \cite{tosicACM}. Our findings here may also help one in interpreting
recent experimental evidence of parallel processing in laminar neocortex
microcircuits \cite{neoc2,neoc3}. That is, a comparison between the model
behavior and experimental results may shed light on the dynamics of these
circuits and their mutual interactions.

We thank I. Erchova, P.L. Garrido and H.J. Kappen for very useful comments.
This work was financed by \textit{FEDER}, \textit{MEyC} and \textit{JA}
under projects FIS2005-00791 and FQM--165. JMC also acknowledges financial
support from the EPSRC-funded COLAMN project Ref. EP/CO 10841/1.

\end{document}